\documentclass[prl,showpacs,twocolumn]{revtex4}
\usepackage{epsfig}
\usepackage{graphicx}
\usepackage{bm}

\newcommand{\rhat}{{\bf \hat{r}}}
\newcommand{\khat}{{\bf \hat{k}}}
\newcommand{\ucheck}{{\bf \check{u}}}
\newcommand{\zerohat}{{\bf \hat{0}}}
\newcommand{\tcheck}{{\bf \check{t}}}

\begin{document}

\title{Anisotropic Random Networks of Semiflexible Polymers}
\author{Panayotis Benetatos and Annette Zippelius}
\affiliation{Institute for Theoretical Physics, Georg-August
  University of G\"ottingen, Friedrich-Hund-Platz 1, 37077
  G\"ottingen, Germany}

\date{\today}

\begin{abstract}
  Motivated by the organization of crosslinked cytoskeletal biopolymers, we present
  a semimicroscopic replica field theory for the formation of
  anisotropic random networks of semiflexible polymers. The networks
  are formed by introducing random permanent crosslinks which fix the
  orientations of the corresponding polymer segments to align with one
  another. Upon increasing the crosslink density, we obtain a
  continuous gelation transition from a fluid phase to a gel where a
  finite fraction of the system gets localized at random positions.
  For sufficiently stiff polymers, this positional localization is
  accompanied by a {\em continuous} isotropic-to-nematic (IN)
  transition occuring at the same crosslink density. As the polymer
  stiffness decreases, the IN transition becomes first order, shifts
  to a higher crosslink density, and is preceeded by an orientational
  glass (statistically isotropic amorphous solid) where the average
  polymer orientations freeze in random directions.

\end{abstract}
\maketitle

\paragraph{Introduction.}

Semiflexible polymers are macromolecules whose behavior is dominated
by their bending stiffness. Recent years have seen an explosion of
interest in them because some of the most important structural
elements of the cytoskeleton (e.g. F-actin) and the extracellular
matrix (e.g. collagen) fall in this category \cite{Kroy_Bausch}. In
vivo, these filaments appear in the form of networks or bundles. A
first step towards understanding the behavior of such complex
aggregates is to study in vitro solutions of many filaments whose
interactions are controled by crosslinkers
\cite{Tempel,Gardel,Wagner}. There is currently significant activity
on the mechanics and elasticity of crosslinked bundles \cite{Clae,HeussBath}. Previous attempts to theoretically
describe the {\it formation} of ordered structures from disordered solutions
use a generalized Onsager approach \cite{Borukhov} or a Flory-Huggins
theory \cite{Zilman_Safran}. In both approaches, the filaments are
modeled as rigid rods and the role of thermal bending fluctuations
(finite persistence length) is neglected.

\begin{figure}
\begin{center}
\leavevmode
\hbox{%
\epsfxsize=1.4in
\epsffile{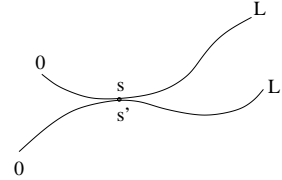}}
\end{center}
\caption{Crosslinker aligning two wormlike chains at arc length $s$ and
  $s'$.\label{crosslink}}
\end{figure}

In this Letter, we consider randomly crosslinked networks of wormlike
chains (WLCs), which are characterized by two parameters: the total
contour length, $L$, and the persistence length, $L_p$ \cite{STY}.
Permanent crosslinks connect randomly chosen pairs of WLCs, such that
parallel alignment of the two chains participating in the crosslink is
enforced (see Fig. \ref{crosslink}). Such crosslinks can be realized
experimentally with short actin-bundling proteins such as fimbrin
\cite{Winder}. The parallel alignment acts like an effective
Maier-Saupe interaction, giving rise to nematic ordering in the gel
with an unexpected strong dependence on the stiffness of the chain. If
the persistence length, $L_p$, of the WLC is sufficiently large,
nematic ordering is observed right at the gel point. The degree of the
ordering transition within the gel fraction is discontinuous as in most nematic
transitions.  However, the orientational ordering is mediated by the
crosslinks, so that only the fraction of chains in the gel exhibit
nematic ordering.  Since the gel fraction goes to zero continuously at
the gel point, the nematic order parameter is also continuous at the
transition. For smaller persistence length, we find a true first order
nematic transition inside the gel phase with the distance from the gel
point increasing with $L/L_p$.
Our model also predicts an orientational glass, which is characterized
by frozen random orientations of the localised chains in the gel. In
this phase the rotational symmetry of the system is broken for any
realization of the crosslinks but is statistically restored.

\paragraph{Model.} We consider a melt of $N$ identical semiflexible
polymers modeled as wormlike chains (WLCs) with contour length $L$ and
persistent length $L_p$ in three-dimensional space. The Hamiltonian of
the system has one part related to the bending stiffness of the WLCs
and another part, $U_{EV}$, ensuring excluded-volume repulsion \cite{ftnt}: 
$$
{\cal H}(\{{\bf r}_i(s)\})=\sum_{i=1}^N\frac{1}{2}{\kappa}\int_0^L ds (\partial_s {\bf
t}_i(s))^2 + U_{{\rm EV}}\;.
$$
Here $\kappa=2L_p k_B T$ denotes the bending stiffness of a WLC and ${\bf
  t}_i(s)= \partial_s {\bf r}_i(s)$ is its tangent vector at arc
length $s\; (0\leq s\leq L)$ with $|{\bf
  t}_i(s)|=1$. 
We introduce $M$ permanent random crosslinks which
constrain the system in such a
way that they fix the positions of the corresponding segments to
overlap  {\em and } their orientations to be {\em parallel} or {\em antiparallel}. The
partition function for a specific configuration of crosslinks reads:
$$
Z({\cal C}_M)=\langle \prod_{e=1}^M \delta ({\bf r}_{i_e}(s_e)-{\bf
  r}_{j_e}(s'_e))\delta ({\bf t}_{i_e}(s_e)-m_e {\bf
  t}_{j_e}(s'_e))\rangle,
$$
where $m_e=\pm 1$ and the average, $\langle...\rangle$, is over all polymer
conformations with the Boltzmann weight $\exp(-\beta {\cal H})$. We
treat the constraints imposed by the crosslinks as quenched disorder,
and our goal is to calculate the disorder-averaged free energy,
$F=-k_B T [\ln Z]$, where $[...]$ denotes averaging over all crosslink
conformations which are determined by the number of crosslinks, $M$,
their positions and their polarities, ${\cal C}_M=\{i_e,j_e ; s_e,
s'_e ;m_e\}_{e=1}^M$.
We assume that a disorder configuration with $M$
crosslinks follows the Deam-Edwards distribution \cite{DE}:
$$
  {\cal P}({\cal C}_M)\propto \frac{1}{M!}
\big(\frac{\mu^2 V}{N}\big)^M\langle\prod_{e=1}^M 
\delta ({\bf r}_{i_e}(s_e)-{\bf
    r}_{j_e}(s'_e))\rangle\;.
$$
The physical content of this distribution is that polymer segments
close to each other in the liquid (uncrosslinked) phase, {\em
  irrespective} of their relative orientation, have a high probability
of getting linked. The parameter $\mu^2$ can be thought of as a
fugacity controlling the mean number of crosslinks per
WLC: $[M]/N$ is of order $\mu^2$.

\begin{figure*}
\begin{minipage}{0.23\textwidth}
\includegraphics*[width=\textwidth]{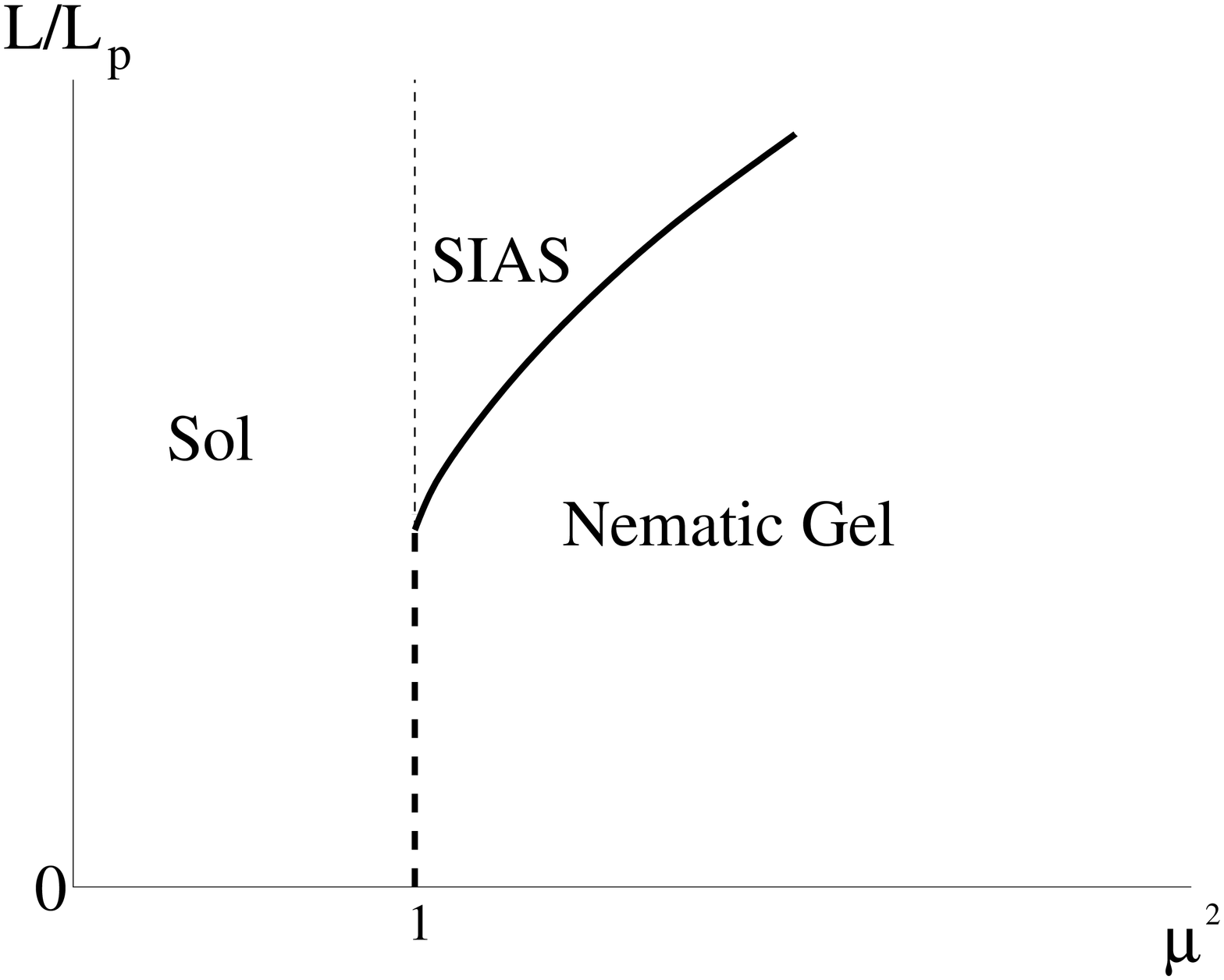}
\end{minipage}
\hspace{1cm}
\begin{minipage}{0.22\textwidth}
\includegraphics*[width=\textwidth]{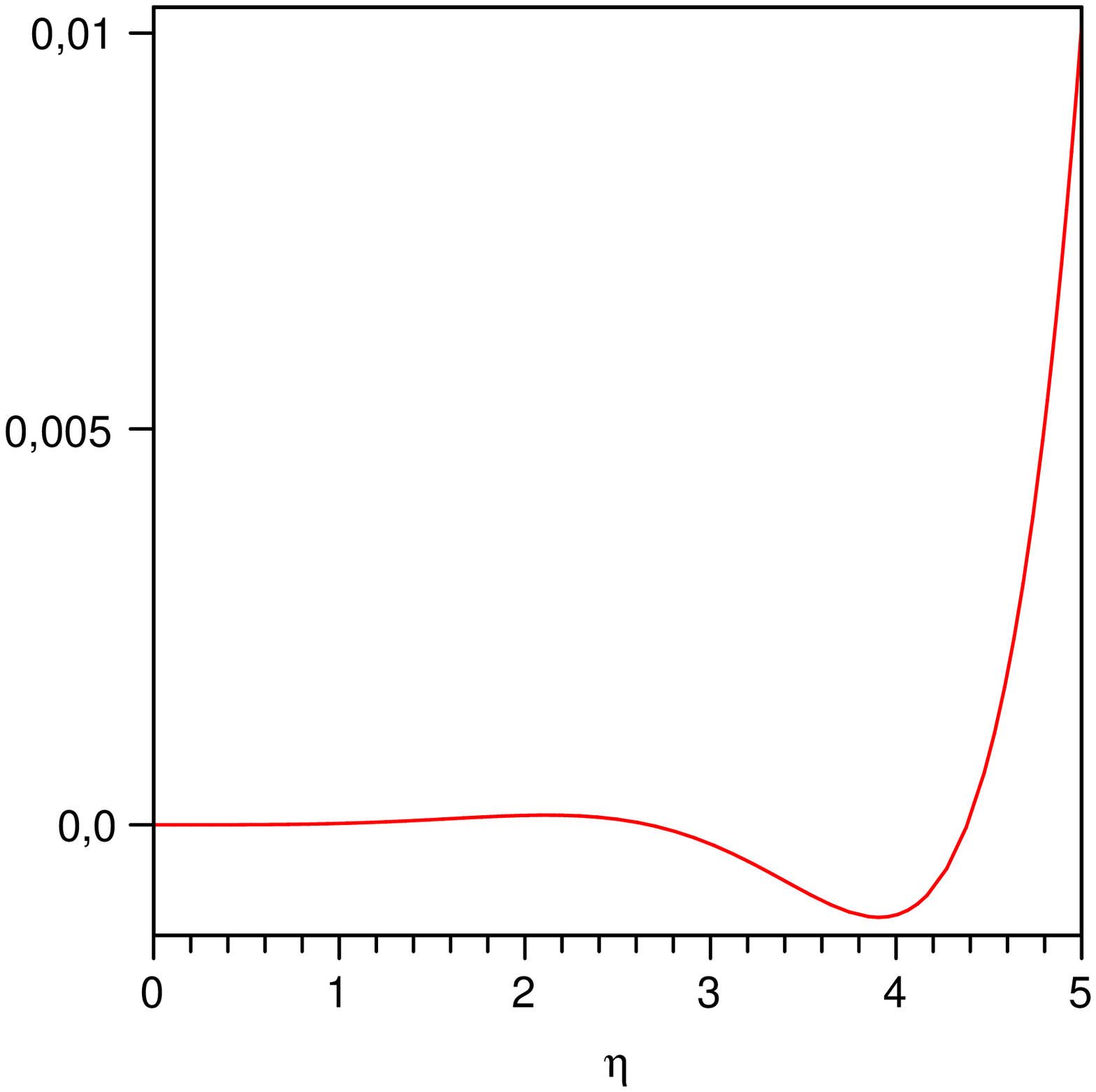}
\end{minipage}
\hspace{0.05cm}
\begin{minipage}{0.22\textwidth}
\includegraphics*[width=\textwidth]{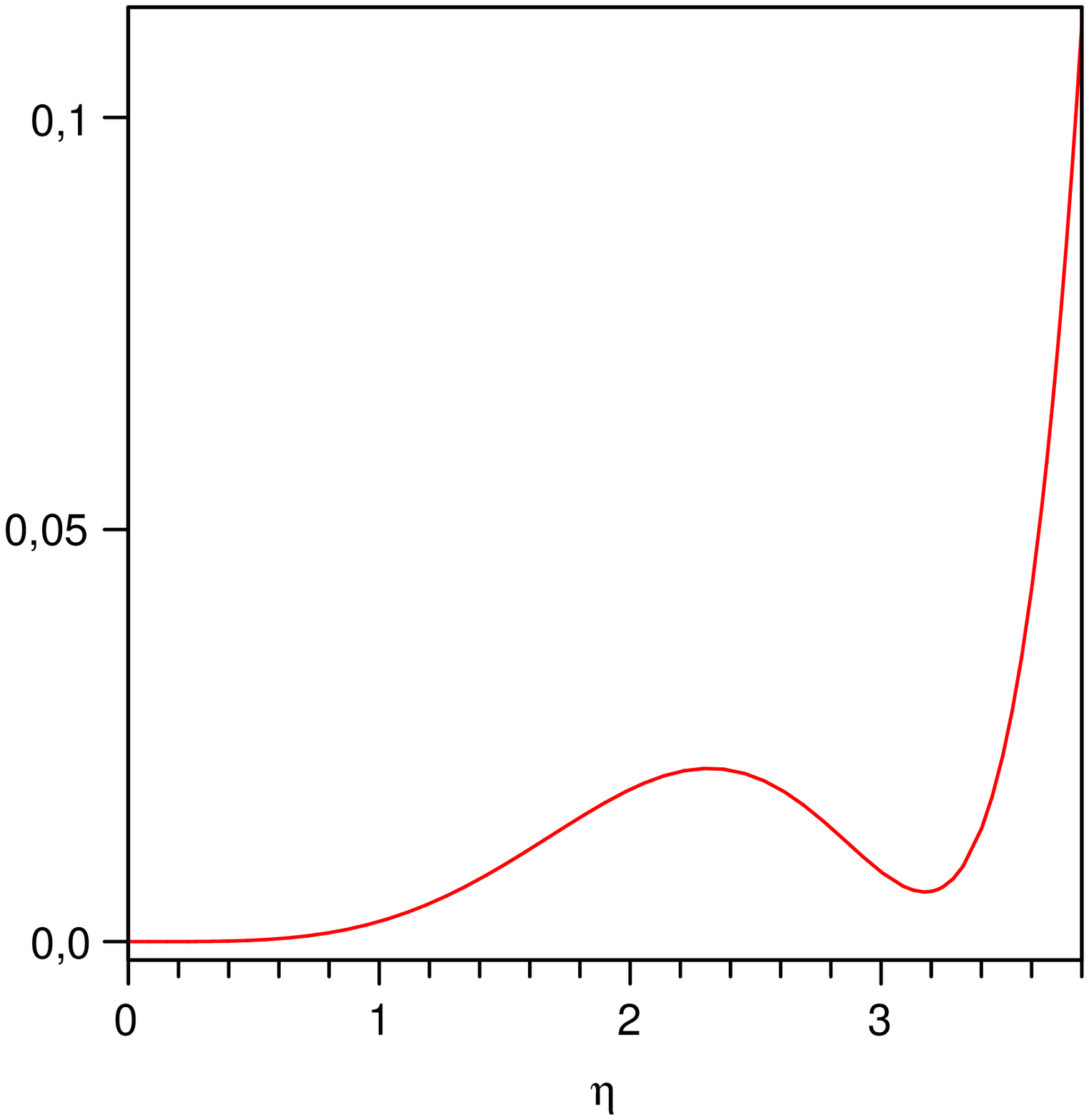}
\end{minipage}
\hspace{0.05cm}
\begin{minipage}{0.22\textwidth}
\includegraphics*[width=\textwidth]{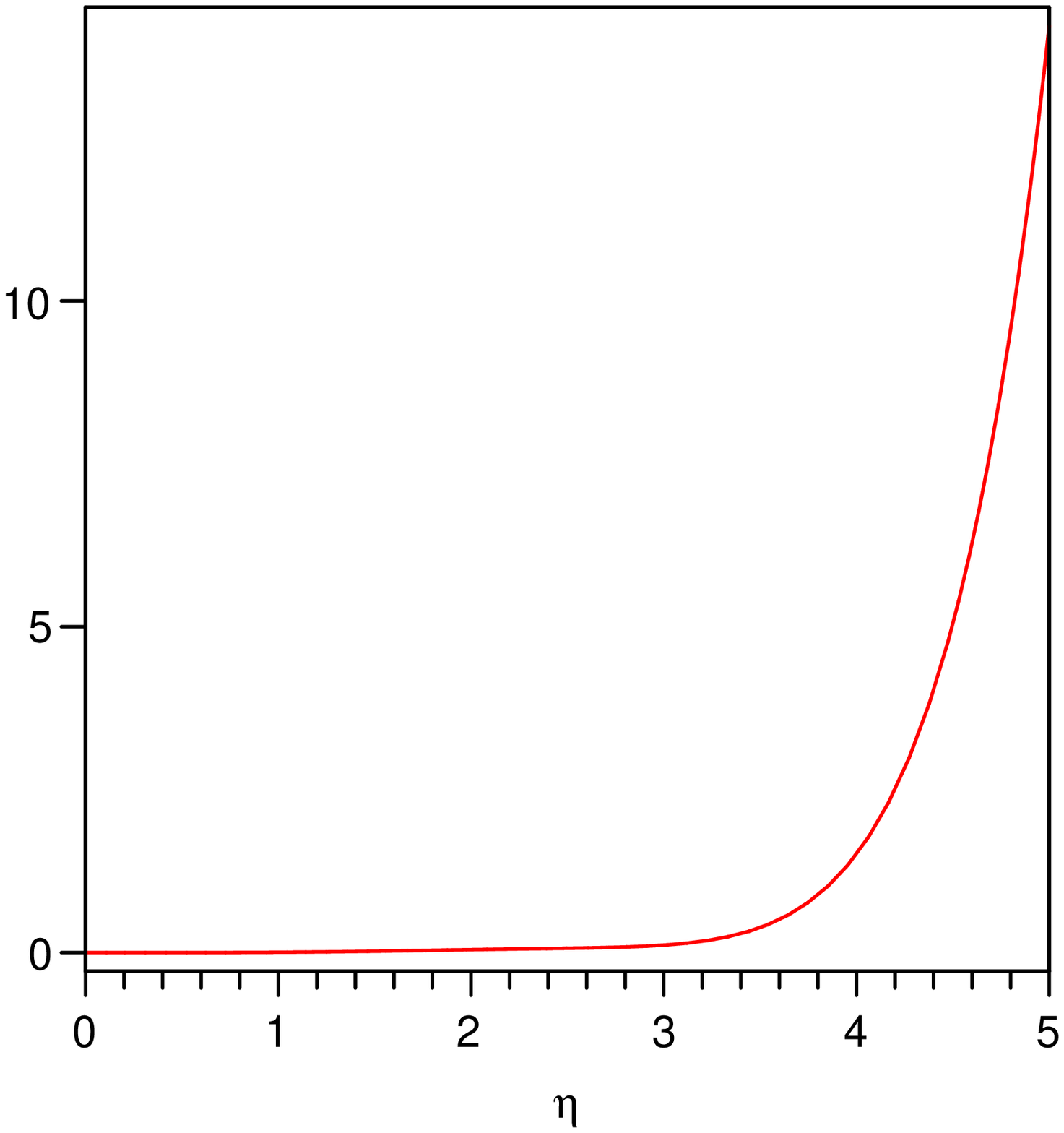}
\end{minipage}
\caption{The first panel shows
  the schematic phase diagram. The other three panels show the free
  energy per WLC (in units of $k_B T$) as a function of $\eta$ for
  $l=0.001$ (second), $l=0.15$ (third) and $l=0.4 $ (fourth) at $\mu^2=\mu_c^2=1$.  \label{free_energy}}
\end{figure*}

\paragraph{Order Parameter and Free Energy.} 
As the number of crosslinks is increased to about one per chain, we
expect a gel transition to an amorphous solid state with a finite
fraction of WLCs localised at random positions. This spatial
localisation due to crosslinking implies for sufficiently stiff chains
a restriction also on the orientation of the chains.
In a phenomenological ansatz, we model the
probability of a single monomer segment, $s$, to be found at position
${\bf x}$ with orientation ${\bf u}$ as
\begin{equation}
\label{pheno}
  <\delta({\bf x} - {\bf r}(s))\delta({\bf
    u}-{\bf t}(s))>\propto e^{-\frac{({\bf x} - {\bf a})^2}{2\xi^2}}
  e^{\eta{\bf u} \cdot {\bf e}}. 
\end{equation}
Here ${\bf a}$ denotes the preferred random position of the monomer,
$\xi$ is the localization length, and ${\bf e}$ is a unit vector with
$\eta$ specifying the degree of orientational ordering. If the nematic
aligning interaction induced by the crosslinks is strong enough, then a nematic
phase arises. In particular for a (uniaxial) nematic gel ${\pm \bf e}$
is the global axis of orientation. If the aligning interaction is weak, then
an orientational glass is expected and ${\bf e}$ is equally likely to
point in any direction.  For a given number of crosslinks, the
effective strength of the nematic interaction is controlled by the
persistence length $L_p$. For large $L_p/L \gg 1$, one crosslink per
chain is enough to approximately fix the orientation of the whole
chain, whereas for $L_p/L \ll 1$ many crosslinks are required to
achieve nematic ordering. Hence we expect a phase diagram as shown in
Fig. (\ref{free_energy}) with $\mu^2$ controlling the number of
crosslinks and $L/L_p$ the polymer flexibility. This phase diagram is
born out by the calculations, as we now sketch.

In order to calculate the disorder averaged free energy, we apply the standard
replica trick, $[\ln Z]=\lim_{n\rightarrow 0}([Z^n]-1)/n$.
We formally decouple individual WLCs from one another by eliminating
the single-polymer degrees of freedom in favor of collective
fields \cite{GCZ}. In the saddle-point approximation, the density of
the system is uniform provided that the excluded-volume interaction is
strong enough to prevent it from collapsing due to the
crosslinks. Note that this interaction cannot induce an
isotropic-nematic transition \'a la Onsager.
The replica free energy per chain then reads:
\begin{equation}
\label{repl_f}
f\{\Omega(\khat,\ucheck)\}=\frac{\mu^2 V}{2}\overline{\sum}_{\khat}
\int_{\ucheck}|\Omega(\khat,\ucheck)|^2-\log{z}\;,
\end{equation}
where
\begin{eqnarray}
\label{partition_function}
& & {z}=\Big\langle\exp\big(\mu^2 V \overline{\sum}_{\khat}
\int_{\ucheck}\Omega(\khat,\ucheck)\nonumber\\ 
& &\frac{1}{2L}\sum_{m=\pm 1}\int_0^L
ds e^{-i\khat \cdot \rhat(s)} \times \delta(\ucheck-m
\tcheck(s))\big)\Big\rangle^{\rm w}_{n+1}\;.
\end{eqnarray}
Here $\khat=({\bf k}^0,{\bf k}^1,...,{\bf k}^n)$, $\ucheck=({\bf
  u}^1,{\bf u}^2,...{\bf u}^n)$, and $\overline{\sum}_{\khat}$ denotes a sum
over all wave-vectors except for $\khat=({\bf 0},{\bf 0},...,{\bf
  0})$. The average is over an $(n+1)$-fold replication of the {\em single}
WLC Hamiltonian. 

In saddle-point approximation, the field $\Omega$
satisfies the self-consistent equation:
\begin{eqnarray}
\label{Omega_sp}
\Omega(\khat,\ucheck)=\sum_{i=1}^N\sum_{m}
\int_0^L
\frac{ds}{2LN} \big\langle e^{-i\khat \cdot \rhat_i(s)}\delta(\ucheck-m
\tcheck_i(s))\big\rangle\;,
\end{eqnarray}
where the average refers to the single chain ``partition function'' of
Eq. (\ref{partition_function}).  $\Omega(\khat,\ucheck)$ acts as an
order parameter which distinguishes between various phases, such as
liquid, crystalline and amorphous solid with or without orientational
order. As discussed above, we focus here on two scenaria for
orientational ordering: nematic gels and statistically isotropic
amorphous solids (SIAS), where the orientation of the chains is frozen
in random directions. Orientational ordering is mediated by the
crosslinks, affecting only chains in the same cluster. Chains in {\it
  finite} clusters, coexisting with the infinite cluster, move and
reorient thermally. Hence only the fraction of localised chains, $Q$,
i.e. those chains which are part of the {\it infinite} cluster,
exhibit orientational order. Therfore we generalise the order
parameter of the isotropic gel to the following form:
\begin{equation}
\label{Order_parameter}
\Omega(\khat,
\ucheck)=(1-Q)\;\delta_{\khat,\zerohat}+Q\;\omega(\khat,\ucheck)\;\delta_{{\bf 0},\Sigma_{\alpha=0}^n{\bf
    k}^{\alpha}}.\nonumber\\
\end{equation}
The gel fraction is denoted by Q and macroscopic translational
invariance requires $\Sigma_{\alpha=0}^n{\bf k}^{\alpha}={\bf 0}$. 

\paragraph{Statistically Isotropic Amorphous Solid.}
In the SIAS the orientation of the WLCs in the gel are frozen, such
that the prefered direction fluctuates randomly from chain to chain.
Rotational symmetry is broken for any single realization of disorder
but is statistically restored. (This type of order has also been
predicted for other systems \cite{EPL,Theissen}). The order parameter
involves an average over all chains, which is equivalent to an average
over all directions. Motivated by the phenomenological picture, we
make the following variational ansatz for the order parameter
\begin{equation}
\label{Omega_sias}
\omega(\khat,
\ucheck)=e^{-\xi^2 \khat^2/2}\, \int d {\bf e}
\prod_{\alpha=1}^n e^{\eta \,{\bf e}\cdot {\bf u}^\alpha}
\big(\frac{\eta}{\sinh(\eta)}\big)^{n}
\end{equation}
where the last factor ensures proper normalization of the order parameter, 
$\int d\ucheck \, \omega(\khat={\bf{\hat 0}},\ucheck)=1$.

Using this ansatz, and taking the limit $n\rightarrow 0$
followed by $V\rightarrow \infty,\,N\rightarrow \infty,\,N/V=const.$,
the free energy per polymer reads:
\begin{eqnarray}
\label{f_sias}
& &f(\xi^2,\eta,Q)=\frac{(Q\mu)^2}{2}
\Big(-\frac{3}{2}\ln(\xi^2)-\frac{1}{54}\eta^4\Big)\nonumber\\
& &-\frac{(Q\mu^2)^2}{2}\Big(-\frac{3}{2}\ln(\xi^2)-
\frac{1}{4}B_1\frac{1}{\xi^2}-\frac{1}{18}B_2\eta^4-
\frac{1}{6}B_3\frac{\eta^2}{\xi^2}\Big)\nonumber\\
& &-\frac{(Q\mu^2)^3}{4}\ln(\xi^2)\;,
\end{eqnarray}
where we have kept only the leading order terms in $Q$, $\xi^{-2}$, and
$\eta$. $B_1$, $B_2$, and $B_3$ depend on $l\equiv L/L_p$ through the
lowest moments of the WLC conformational probability
distribution. 

Stationarity of the free energy with respect to $Q$,
$\xi$, and $\eta$ yields a continuous gelation transition at $\mu_c^2=1$
characterized by a nonzero gel fraction $Q\sim 2(\mu^2-1)$,
independent of $l$. These results are universal for the gelation
transition in the saddle point approximation and have been confirmed
for various models (e.g. \cite{GCZ,Huthmann,Theissen}).
The localization length and the degree of
orientational order depend on the stiffness of the WLC.
In the flexible limit, the localisation length is determined by the
radius of gyration $\xi^2\sim (\mu^2-1) L L_p$ and the orientational order goes
to zero as $\eta^2 \sim L_p/L$. In the stiff rod limit ($\l
\rightarrow 0$) $\eta^2 \xi^2$ diverges at $\mu^2=1$. Our perturbative
approach breaks down in this singular case where a single increasingly
long stiff rod is formed.

\paragraph{Nematic Gel.} The alignment of the crosslinked polymer
segments may trigger an isotropic-to-nematic phase transition,
provided a macroscopic cluster of crosslinked polymers exists. 
For a uniaxial nematic the two directions $\pm {\bf e}$ are
equivalent. Hence we sum over these two directions $\pm{\bf e}$ in the
pheneomenological ansatz (\ref{pheno}) and generalize the order
parameter to include nematic ordering:
\begin{eqnarray}
\label{Omega_nem}
\omega(\khat,
\ucheck)=e^{-\xi^2\khat^2/2}\big(\frac{\eta}{\sinh{\eta}}\big)^n
\prod_{\alpha=1}^n  \cosh{(\eta{\bf u^{\alpha}}  \cdot {\bf e})}. 
\end{eqnarray}

The experimentally accessible nematic order parameter, 
\begin{eqnarray}
\label{S_discr}
{\cal S}=\frac{1}{N}\sum_{i=1}^N\frac{1}{2L}\int_0^L ds \langle 3({\bf
  e}\cdot {\bf t}_i(s))^2-1\rangle\;,
\end{eqnarray}
is obtained from the generalised order parameter field as
\begin{eqnarray}
\label{S_def}
{\cal S}= \frac{Q\, \eta}{4}\int_{-1}^{+1} dx
\frac{\cosh{(\eta x)}}{\sinh(\eta)}(3\cos^2x-1)\;,
\end{eqnarray}
showing clearly that only the gel fraction contributes to the nematic
ordering. Even if the nematic transition is of first order - as it
will turn out - the jump in the nematic order parameter may be very
small due to a small gel fraction and in fact ${\cal S}$ may even be
continuous (see below).

If we substitute the above ansatz, Eq. (\ref{Omega_nem}), into the
saddle-point free energy, we obtain:
\begin{equation}
\label{f_nematic}
f(\xi^2,\eta,Q)=f_{\rm p}(\xi^2,Q)+f_{\rm n}(\eta,Q)+{\cal O}
(\frac{Q^2 \eta^2}{\xi^4})\;,
\end{equation}
The positional part of the free energy, $f_{\rm p}(\xi^2,Q)$, is
exactly the same as that obtained from Eq. (\ref{f_sias}) by setting
$\eta=0$, whereas the lowest-order in $Q$ orientational part reads:
\begin{widetext}
\begin{eqnarray}
\label{f_n}
f_{\rm  n}(\eta,Q)=\frac{(Q\mu)^2}{2}\ln\Big\{\eta\frac{\cosh(\eta)\sinh(\eta)+\eta}
{2\sinh^2(\eta)}\Big\}
-\frac{(Q\mu^2)^2}{l^2}\int_0^L ds\int_0^s ds'\ln\Big\{\frac{\langle\cosh(\eta \:{\bf
    e}\cdot{\bf t}(s))\cosh(\eta \:{\bf
    e}\cdot{\bf t}(s'))\rangle_{\rm w}\eta^2}{\sinh^2(\eta)}\Big\}\;,
\end{eqnarray}
\end{widetext}
where $\langle...\rangle_{\rm w}$ denotes averaging over the WLC
conformations.
For finite $\eta$, stationarity of the free energy with
respect to variations in $Q$ and $\xi^2$ yields exactly the same
results as in the gelation transition considered in the previous
paragraph. Since $\xi^{-2}\sim Q$, the term which couples positional
and orientational localization is of higher order and can be neglected
close to the gelation transition ($Q\ll 1$). 


If we expand the nematic part of the free energy up to quartic order
in $\eta^2$, we obtain the typical Landau free energy which yields a
first order phase transition from an isotropic phase with $\eta=0$ to
a nematic phase with finite $\eta$:
\begin{eqnarray}
\label{Landau_3}
  \frac{1}{Q^2}f_{\rm n}(\eta)=a(l, \mu^2)\eta^4+b(l,
  \mu^2)\eta^6+c(l, \mu^2)\eta^8, 
\end{eqnarray}
with $b(l, \mu^2)<0$. The coefficients depend on the control
parameters of the system which are the crosslink density (through
$\mu^2$) and the single WLC flexibility (through $L/L_p$).  The free
energy (\ref{Landau_3}) is shown for $\mu^2=1$ and various values of
$l$ in Fig. (\ref{free_energy}). For stiff chains ($l=0.001$), the
global minimum occurs at finite $\eta$ already at the gel point,
whereas for more flexible chains ($l=0.15$), the minimum at finite
$\eta$ represents a metastable state with the global minimum at
$\eta=0$. In this case, the transition to the nematic happens at a
higher crosslink density: $\mu_*^2=1+{\cal O}(l)$. Finally, rather
semiflexible chains ($l=0.4$ in Fig.  \ref{free_energy}) exhibit not
even a metastable nematic state at $\mu^2=1$, but do so at higher
croslink densities: a nematic metastable minimum appears at $\mu_1^2
>1$, which becomes global at the transition point $\mu_*^2 >\mu_1^2$.
The resulting phase diagram is shown schematically in
Fig. (\ref{free_energy}). 

Since the degree of
orientational order, $\eta$, is finite at the transition, one cannot
make quantitative predictions based on the Landau expansion of the
free energy. However, we can improve it by expanding
$\langle...\rangle_{\rm w}$ in Eq. (\ref{f_n}) in Legendre
polynomials.
The truncation of the power series expansion becomes less reliable as
the WLCs become stiffer.
We have checked the behavior of the free energy keeping terms up to
$\eta^{14}$, and the {\em qualitative} features of its dependence on
the crosslink density are robust.
As we increase the crosslink density
beyond $\mu_*^2$, there is a value at which $f_{\rm n}(\eta)$ becomes
unstable for large $\eta$. At that point, our perturbative approach
collapses, since it is based on the assumption of the
positional-orientational decoupling close to the gelation transition
which is valid for finite $\eta$. Also, at the stiff rod limit
($L/L_p\rightarrow 0$), $f_{\rm n}(\eta)$ becomes unstable at
$\mu^2=1$.

\paragraph{Conclusions.}
In this Letter, we have shown how the geometry of the crosslinks together
with the stiffness of the constitutent chains control the
orientational order of the random macromolecular network which is
formed upon gelation. Whereas the gel formation is solely controlled
by the number of crosslinks, the orientational order induced by
localisation is sensitive to both the number of crosslinks and the
stiffness of the WLCs. Rather stiff chains tend to exhibit nematic order 
right at the gel point with a {\it continuous} isotropic to nematic
transition. More flexible chains exist in a phase
with frozen random orientation right at the gel point and show a first
order nematic transition only at higher crosslink density. 
Future investigations may extend this work to consider
different crosslinking geometries; one important example is polar
ordering induced by crosslinks. It would also be of interest to study
the possibility of combined nematic and SIAS orderings. 

We thank Erwin Frey for useful discussions and acknowledge financial
support from the DFG under grant ZI 209/7.


\begin{thebibliography}{}
\bibitem{ftnt}
$U_{EV}$ does not imply any finite-size hard-core repulsion.
\end{thebibliography}
\end{document}